\documentclass[aps,prb,twocolumn,showpacs,groupedaddress,10pt]{revtex4-1}

\usepackage{graphicx}
\usepackage{amsmath}
\usepackage{verbatim,dsfont}

\begin{document}

\title{RKKY interaction in a disordered two-dimensional electron gas \\ with Rashba and Dresselhaus spin-orbit couplings}

\author{Stefano Chesi and Daniel Loss}
\affiliation{Department of Physics, University of Basel, Klingelbergstrasse 82, 4056 Basel, Switzerland}
\date{\today}

\begin{abstract}
We study theoretically the statistical properties of the Ruderman-Kittel-Kasuya-Yosida (RKKY) interaction between localized magnetic moments in a disordered two-dimensional electron gas with both Rashba and Dresselhaus spin-orbit couplings. Averaging over disorder, the static spin susceptibility tensor is evaluated diagrammatically in the mesoscopic (phase-coherent) regime. The disorder-averaged  susceptibility leads to a twisted exchange interaction suppressed exponentially with distance, whereas the second-order correlations, which determine the fluctuations (variance) of the RKKY energy, decay with the same power-law as in the clean case. We obtain analytic expressions in the limits of large/small spin orbit interactions and for equal Rashba and Dresselhaus couplings. Beside these limiting cases, we study numerically the variance of the RKKY interaction in the presence of pure Rashba spin-orbit coupling. Our results are relevant for magnetic impurities or nuclear moments embedded in III-V two-dimensional heterostructures or in contact with surface states of metals and metal alloys, which can display a sizable Rashba spin-orbit coupling. 
\end{abstract}




\maketitle

\section{Introduction}

The Ruderman-Kittel-Kasuya-Yosida (RKKY) magnetic interaction between nuclei \cite{Ruderman1954} or magnetic impurities \cite{Kasuya1956,Yosida1957} in bulk metals is determined by the nonlocal spin susceptibility of the conduction electrons. A given localized magnetic moment in contact with a clean electron gas at low temperature produces an electronic spin polarization which is oscillatory (with period half of the Fermi wavelength) and decays in magnitude as a power law of the distance from the impurity. Other distant magnetic moments interact with such polarization pattern. In the presence of weak non-magnetic disorder for the electron gas, the configuration average of such long range interaction is exponentially suppressed beyond the mean-free path.\cite{DeGennes1962} This is due to the randomization of the oscillatory tail of the interaction but does not imply that the interaction strength is exponentially suppressed. The magnetic interaction is still long-range and the variance is found to decay with the same power law of the clean case.\cite{Zyuzin1986,Bulaevskii1986}

Similar results are also valid for electrons confined in two dimensions.\cite{Bergmann1987,Jagannathan1988,Jagannathan1992} Furthermore, in semiconductor heterostructures, the long range nature of the RKKY interaction is enhanced in the presence of nonanalytic corrections to the spin susceptibility,\cite{Simon2007,Simon2008,Chesi2009,Zak2010} and a magnetic ordering transition of the nuclear spin system was predicted.\cite{Simon2007,Simon2008} In the following, we neglect these interaction effects (becoming relevant for low electron densities) and focus instead on the role of the spin-orbit coupling, which is responsible of anisotropic magnetic interactions. Spin-orbit scattering by nonmagnetic impurities is important to determine the anisotropy field of bulk spin glasses,\cite{Levy1981} and can be treated by standard diagrammatic techniques.\cite{Rammer1998,Zyuzin1986,Bergmann1987,Jagannathan1988,Jagannathan1992} Its effect on the statistical properties of the magnetic interactions in two dimensions was studied in Refs.~\onlinecite{Jagannathan1988,Jagannathan1992}, which in particular obtain a crossover  from Heisenberg to Ising behavior.

In this paper we assume that spin-orbit scattering from the impurities is not present, but we consider translationally invariant spin-orbit interactions due to the band structure.\cite{WinklerSpringer03} The effect of the Rahsba spin-orbit coupling\cite{Bychkov1984} was examined in Refs.~\onlinecite{Imamura2004,Huang2006,Lyu2007,Lai2009} and a simple form of a twisted RKKY exchange interaction was derived at large distance.\cite{Imamura2004} A discussion including also the Dresselhaus spin-orbit interaction\cite{Dresselhaus1955} can be found in Ref.~\onlinecite{Mross2009}. All such studies neglect the effect of disorder but this assumption is not always justified, since the spin-orbit length can be larger than the mean-free path in realistic conditions. It is the main purpose of this paper to examine in detail the interplay of disorder with Rashba and Dresselhaus spin-orbit couplings.

Besides being applicable to III-V heterostructures, where the electron gas mediates interactions between nuclear spins, our results are relevant for isolated magnetic impurities on metal surfaces. In recent experiments, the RKKY interaction of a pair of such adatoms could be directly probed as a function of their relative position.\cite{Wahl2007,Meier2010} On the other hand, surface states with a sizable Rashba spin-orbit coupling exist, as first observed on Au(111).\cite{LaShell1996,Hoesch2004} Much larger spin-orbit splittings are found in metal alloys\cite{Pacile2006,Ast2007} and several methods exist to modify the strength of the spin-orbit coupling in these systems.\cite{Forster2003,Cercellier2006,Bentmann2009} 
Finally, Rashba and Dresselhaus spin-orbit coupling could influence the properties of low-dimensional magnetic semiconductors as (In,Mn)As and (Ga,Mn)As where the RKKY interaction is mediated by holes.\cite{Dietl1997,Ohno2000,Chiba2008,Burch2008,Richardella2010} Indeed, a large spin-orbit interaction is present in two-dimensional heterostructures for holes,\cite{WinklerSpringer03,Winkler2000,Winkler2002} but the dominant spin-orbit coupling is in this case cubic in momentum instead of being linear.\cite{WinklerSpringer03, Winkler2000, Chesi2007} Although our results are not directly applicable, it would be possible to treat this relevant case in a similar fashion.

The paper is organized as follows: In Sec. \ref{sec_general} we discuss the main definitions and the general formalism of our work. The diagrammatic expressions for the disorder-averaged quantities of interest are introduced. An explicit calculation of the susceptibility tensor can be found in Sec.~\ref{sec_average} and the second order-correlations of the susceptibility tensor are obtained in Sec~\ref{sec_limits_correlations} for some important limiting cases. More generally, the variance of the RKKY interaction has to be evaluated numerically and the calculation is presented with many details in Sec.~\ref{sec_rashba_only} for pure Rashba spin-orbit coupling. Finally, Sec.~\ref{the_end} contains our conclusions and a number of technical points are discussed in Appendices \ref{appendix_cooperon_fluctuations}-\ref{eta_simmetries}.

\section{General Formalism}\label{sec_general}

We consider a system of localized moments (e.g. magnetic impurities or nuclear spins) interacting with a free electron gas. Restricting ourselves to a specific pair of such impurities located at positions ${\bf R}_1$, ${\bf R}_2$ with spin operators ${\bf I}_1$, ${\bf I}_2$ we have
\begin{equation}\label{H0_contact}
H=\sum_i H_{el}(i)+J \sum_{j=1,2} {\bf S}({\bf R}_j)\cdot {\bf I}_j ,
\end{equation}
where $H_{el}(i)$ is the single-particle electron Hamiltonian and the second term in Eq.~(\ref{H0_contact}) is a contact interaction with the two magnetic impurities, expressed in terms of the electron spin density (in units of $\hbar/2$)
\begin{equation}\label{contact_int}
{\bf S}({\bf R}) = \sum_i \delta({\bf r}_i-{\bf R} ) \,  \boldsymbol{\sigma}_i,
\end{equation}
with ${\bf r}_i$, $\boldsymbol{\sigma}_i$ the position and Pauli spin operators of electron $i$. We consider here point-like magnetic impurities, i.e., with spatial extent much smaller than the Fermi wavelength.\cite{Smirnov2009} We also assume an isotropic interaction of the Heisenberg type, e.g., describing the hyperfine interaction of conduction electrons in III-V semiconductors, but the following discussion can also be easily adapted to an Ising-like interaction, more appropriate for holes.\cite{Fischer2008} 
 
In a coordinate system with $x,y$ respectively along the crystal axes $[100]$ and $[010]$, the single-particle electron Hamiltonian has the following form
\begin{equation}\label{so_hamilton}
H_{el}=\frac{{\bf p}^2}{2m}+\alpha (p_y \sigma_x-p_x \sigma_y)+\beta (p_x \sigma_x-p_y \sigma_y) +V({\bf r}),
\end{equation}
which includes the Rashba and Dresselhaus spin-orbit couplings and the disorder potential $V({\bf r})$. Equation~(\ref{so_hamilton}) has a simpler form if $x, y$ are along $[110]$ and $[\bar 1 1 0]$, as assumed through the rest of the paper
\begin{equation}\label{so_hamilton2}
H_{el}=\frac{{\bf p}^2}{2m}+\alpha_- p_y \sigma_x - \alpha_+ p_x \sigma_y +V({\bf r}),
\end{equation}
where $\alpha_\pm = \alpha \pm \beta$ define the following spin-orbit lengths
\begin{equation}
\lambda_\pm = \frac{1}{m|\alpha_\pm|}.
\end{equation}
Notice that Eqs.~(\ref{so_hamilton}) and (\ref{so_hamilton2}) have the same form if $\beta=0$, since the Rashba spin-orbit interaction is rotationally invariant in the $xy$-plane. In this case, a single spin-orbit length $\lambda=1/m|\alpha|$ appears. 

By making use of second-order perturbation theory in $J$, the effective RKKY interaction between ${\bf I}_1$ and ${\bf I}_2$ is obtained as follows\cite{Ruderman1954, Simon2008}
\begin{equation}\label{Himp_definition}
H_{12}=-J^2 \sum_{i,j} I_{1i}I_{2j} \chi_{ij}({\bf R}_1,{\bf R}_2),
\end{equation}
in terms of the static spin susceptibility tensor of the electron system $\chi_{ij}({\bf R}_1,{\bf R}_2)$, given by\cite{TheBook}
\begin{equation}\label{chidef_real_omega}
\chi_{ij}({\bf R}_1,{\bf R}_2)=-\frac{i}{\hbar}\int_0^\infty \langle [S_i({\bf R}_1,t), S_j({\bf R}_2)]_- \rangle e^{-\eta t} dt,
\end{equation}
where $\langle \ldots \rangle$ denotes a thermal average, $[a,b ]_\pm=ab \pm ba$, $\eta=0^+$, and
${\bf S}({\bf R},t)=e^{-\frac{i}{\hbar}\sum_j H_{el}(j)t}{\bf S}({\bf R}) e^{\frac{i}{\hbar}\sum_j H_{el}(j)t}$. As in Refs.~\onlinecite{Zyuzin1986,Jagannathan1988}, it is convenient to use the Matsubara technique and by analytic continuation to imaginary frequencies Eq.~(\ref{chidef_real_omega}) is rewritten 
\begin{equation}\label{chidef}
\chi_{ij}({\bf R}_1,{\bf R}_2)=-T \sum_{n} {\rm Tr}[\sigma_i G_{\omega_n}({\bf R}_1,{\bf R}_2) \sigma_j G_{\omega_n}({\bf R}_2,{\bf R}_1)].
\end{equation}
In Eq.~(\ref{chidef}), $T$ is the temperature, $\omega_n=(2n+1)\pi T$ are Matsubara frequencies (we set $k_B=\hbar=1$ in the following), the trace is taken over the spin indexes, 
and $G_{\omega}({\bf R}_1,{\bf R}_2)$ is the single particle electron Green's function
\begin{equation}
G_{\omega}({\bf R}_1,{\bf R}_2)=\langle {\bf R}_1 |(i\omega-H_{el}+\epsilon_F)^{-1} | {\bf R}_2 \rangle ,
\end{equation}
where $\epsilon_F$ is the Fermi energy and $|\bf R \rangle$ are eigenstates of the position operator.

Since the disorder potential $V({\bf r})$ is unknown, we perform an average over its possible realizations (denoted by an overbar). We introduce the following notations for the susceptibility tensor
\begin{equation}\label{chi_av}
\overline{\chi_{ij}}({\bf R})=\overline{\chi_{ij}({\bf R}_1,{\bf R}_2)},
\end{equation}
and its second-order correlations
\begin{equation}\label{chi_corr}
\overline{\chi_{ij}\chi_{i'j'}}({\bf R})= \overline{\chi_{ij}({\bf R}_1,{\bf R}_2) \chi_{i'j'}({\bf R}_1,{\bf R}_2) },
\end{equation}
where we could set ${\bf R}_1= {\bf 0}$ and ${\bf R}_2= {\bf R} =  R\cos\varphi \, {\bf e}_x + R\sin\varphi \, {\bf e}_y$, since translational invariance is restored. Here, ${\bf e}_i$ are unit vectors along the coordinate axes.

By assuming the simplest case of $\delta$-correlated disorder $\overline{V({\bf r})V({\bf r}')}=(m\tau)^{-1}\delta({\bf r}-{\bf r}')$, Eqs.~(\ref{chi_av}) and (\ref{chi_corr}) can be calculated to leading order in $1/k_{F}\ell$, where $k_{F}=m v_F=\sqrt{2 m \epsilon_F} $ is the Fermi momentum and $\ell=v_F \tau $ is the mean-free path. The standard diagrammatic technique can be applied by including the spin-orbit coupling in the disorder-averaged Green's function $G_{\omega}({\bf R})= \overline{G_{\omega}({\bf R}_1,{\bf R}_2)}$. In the self-consistent Born approximation $G_{\omega}({\bf R})$ is given by
\begin{equation}\label{Fourier_transf_G}
G_{\omega}({\bf R})=\int \frac{d{\bf k}}{(2\pi)^2} \, G_{\omega}({\bf k}) e^{-i {\bf k}\cdot {\bf R}},
\end{equation}
where\cite{Chalaev2005,Duckheim2006,Chalaev2009}
\begin{equation}\label{G_k}
G_{\omega}({\bf k})=  
\frac{i\left(\omega + \frac{{\rm sgn}  \, \omega}{2\tau}\right)-\frac{{\bf k}^2}{2m}+\epsilon_F+\alpha_- k_y \sigma_x-\alpha_+ k_x \sigma_y}{\left[i\left(\omega + \frac{{\rm sgn} \, \omega}{2\tau}\right) -\frac{{\bf k}^2}{2m}+\epsilon_F\right]^2  -\alpha^2_- k_y^2-\alpha^2_+ k_x^2}.
\end{equation}
We refer to Sec.~\ref{sec_average} for an explicit evaluation of Eq.~(\ref{Fourier_transf_G}) in the asymptotic limit $R \gg 1/k_F$. 

\subsection{General formulas for $\overline{\chi_{ij}}$ and  $\overline{\chi_{ij}\chi_{i'j'}}$}

\begin{figure}
\includegraphics[width=0.21\textwidth]{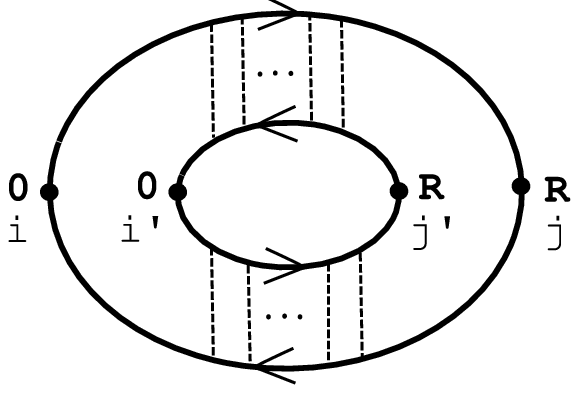}\hspace{0.5cm}
\includegraphics[width=0.21\textwidth]{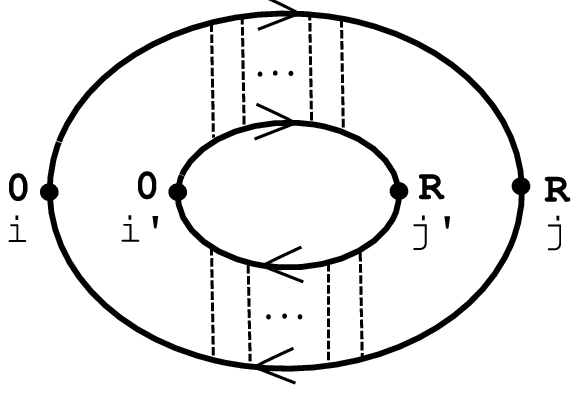}
\caption{\label{diagrams_variance} Diffuson (left) and cooperon (right) diagrams contributing to the second-order correlations of the RKKY interaction tensor given in Eq.~(\ref{correlations_formula}).}
\end{figure}

The leading contribution to Eq.~(\ref{chi_av}) has a simple form:
\begin{equation}\label{chi_leading}
\overline{\chi_{ij}}({\bf R})\simeq -T \sum_{n} {\rm Tr}[\sigma_i G_{\omega_n}({\bf R}) \sigma_j G_{\omega_n}(-{\bf R})] ,
\end{equation}
which corresponds to a single empty bubble diagram. This expression applies in the diffusive limit, when $R\gg \ell$ and vertex corrections can be neglected. On the other hand, $\overline{\chi_{ij}\chi_{i'j'}}({\bf R})$ is obtained to leading order as a sum of two series of diagrams, involving either diffuson or cooperon propagators (see Fig.~\ref{diagrams_variance}).\cite{Zyuzin1986,Bergmann1987,Jagannathan1988} The equality of diffuson and cooperon diagrams for vanishing
external magnetic field is known for other fluctuation calculations (in particular, for the universal conductance fluctuations\cite{Rammer1998}) and we show in Appendix \ref{appendix_cooperon_fluctuations} that this occurs also here. Therefore, we can write the final result in terms of the diffuson propagator $D_{\bf R}(\omega)$ as follows
%
%
\begin{eqnarray}\label{correlations_formula}
\overline{\chi_{ij}\chi_{i'j'}}({\bf R}) =  \frac{(m\tau)^4}{\pi^2}\int_0^\infty \omega d\omega \sum_{\mu,\nu, \mu',\nu'} D^{\mu\nu}_{\bf R}(\omega)D^{\mu'\nu'}_{-{\bf R}}(\omega) \nonumber \\
\times {\rm Tr}(\sigma_i \sigma_\mu\sigma_{i'}\sigma_{\nu'}){\rm Tr}(\sigma_j \sigma_{\mu'}\sigma_{j'}\sigma_{\nu}). \quad
\end{eqnarray}
The diffuson $D_{\bf R}(\omega)$ is a $4\times 4$ matrix given by
\begin{equation}\label{diffuson_def}
D_{\bf R}(\omega)= \frac{1}{2m\tau}  \int \frac{d{\bf q}}{(2\pi)^2}  e^{-i {\bf q}\cdot {\bf R}} [\mathds{1}-X_{D}({\bf q},\omega)]^{-1},
\end{equation} 
with
\begin{equation}\label{XD_def}
X_{D}^{\mu\nu}({\bf q},\omega)=\frac{1}{2m\tau}\int \frac{d{\bf k}}{(2\pi)^2} {\rm Tr}[\sigma_\mu G_{\frac{\omega}{2}}({\bf k}) \sigma_\nu G_{-\frac{\omega}{2}}({\bf k}-{\bf q})],
\end{equation}
where the indices $\mu, \nu$ assume values $0,x,y,z$ ($\sigma_0$ is the identity and $\sigma_{x,y,z}$ are the Pauli matrices). In the diffusive limit $ q\ell, |\omega| \tau \ll 1$ and for small spin-orbit couplings $\lambda_\pm \gg \ell$,  $X_D({\bf q},\omega)$ is given by\cite{Chalaev2005,Duckheim2006,Chalaev2009}
\begin{eqnarray}\label{X_D_result}
&&X_D({\bf q}, \omega) =\left[1-\frac{(q\ell)^2}{2}-|\omega|\tau \right]\mathds{1} \nonumber \\
&& -2 m \ell^2  \left(
\begin{array}{cccc}
0 & 0 & 0& 0 \\
0 & m \alpha_+^2 & 0 & i \alpha_+ q_x \\
0 & 0 & m \alpha_-^2 & i \alpha_- q_y\\
0 & -i \alpha_+ q_x& -i \alpha_- q_y & m (\alpha_+^2 + \alpha_-^2)\\
\end{array}
\right), \quad 
\end{eqnarray}
which allows to evaluate Eqs.~(\ref{diffuson_def}) and (\ref{correlations_formula}). 

\section{Asymptotic expression of the susceptibility tensor}\label{sec_average}

We first study the disorder-averaged susceptibility tensor, given by Eq.~(\ref{chi_leading}). Clearly, $\overline{\chi_{ij}}({\bf R})$ is immediately obtained from the real space representation of the Green's function [see Eqs.~(\ref{Fourier_transf_G}) and (\ref{G_k})] and we start from the angular integral
\begin{equation}
G_{\omega}({\bf R})=\int_0^\infty \frac{k dk }{(2\pi)^2}\int_0^{2\pi} d\varphi_p \, G_{\omega}({\bf k}) e^{-i k R \cos(\varphi_p - \varphi)},
\end{equation}
which can be performed with the method of the stationary phase if  $kR \gg 1$. For small spin-orbit couplings, this condition is realized when $R\gg 1/k_F$ because the largest contribution to the integral is from the region of the Fermi surfaces. A sum of two terms is obtained, corresponding to $\varphi_p \simeq \varphi$ and $\varphi_p \simeq \varphi+\pi$ ($\varphi$ is the azimuthal angle of $\bf R$)
\begin{eqnarray}
G_{\omega}({\bf R}) \simeq \sum_{\pm} \int_0^\infty \frac{dk }{(2\pi)^2}  \,   \sqrt{\frac{2\pi k}{ R}} \, e^{\mp i(kR-\pi/4)}\quad
 \nonumber\\
\times \frac{i\left(\omega + \frac{{\rm sgn} \, \omega}{2\tau}\right)-\frac{k^2}{2m}+\epsilon_F \pm k \boldsymbol{\alpha}_\varphi \cdot \boldsymbol{\sigma}}{\left[i\left(\omega + \frac{{\rm sgn} \, \omega}{2\tau}\right) -\frac{k^2}{2m}+\epsilon_F\right]^2  - k^2 \alpha_\varphi^2},
\end{eqnarray}
where $\boldsymbol{\alpha}_\varphi=\alpha_- \sin\varphi \, {\bf e}_x -\alpha_+ \cos\varphi \, {\bf e}_y$. The main contribution to the $k$-integrals is from the poles located (to leading order in the spin-orbit coupling) at $k_\pm =k_F \pm m \alpha_\varphi +i\left(\omega + \frac{{\rm sgn} \, \omega}{2\tau}\right)/v_F$. If $\omega>0$, only the term of the sum proportional to $e^{+ i(kR-\pi/4)}$ gives a nonvanishing contribution
\begin{equation}\label{G_planewaves}
G_{\omega>0}({\bf R}) \simeq     \frac{-i m}{\sqrt{ 2\pi k_F R}} \sum_\pm e^{i(k_\pm R-\pi/4)}
\frac12 \left( 1 \pm \frac{\boldsymbol{\alpha}_\varphi \cdot \boldsymbol{\sigma}}{\alpha_\varphi}\right).
\end{equation}
The physical meaning of this result is rather transparent since $\operatorname{Re}[k_\pm]$ are the Fermi momenta of the two spin branches along the direction of $\bf R$. Furthermore, Eq.~(\ref{G_planewaves}) involves the spin projectors on the eigenstates of $\boldsymbol{\alpha}_\varphi \cdot \boldsymbol{\sigma}$, determined by the spin-orbit coupling of Eq.~(\ref{so_hamilton2}) when ${\bf k}$ is along ${\bf R}$. By combining this result with the expression at $\omega<0$ we find
\setlength\arraycolsep{1pt}
\begin{eqnarray}\label{Gclean_largeR_RD}
G_{\omega }({\bf R})\simeq &-& i \frac{ ({\rm sgn}\, \omega)  \, m }{\sqrt{2\pi k_F R}} \,  e^{i \,({\rm sgn}\, \omega) \,   (k_F R-\pi/4)} \, e^{-(|\omega|+1/2\tau) R/v_F} \nonumber \\
&\times & \left[\cos (m \alpha_\varphi R) + i \frac{\boldsymbol{\alpha}_\varphi \cdot \boldsymbol{\sigma}}{\alpha_\varphi} \sin (m\alpha_\varphi R) \right],
\end{eqnarray}
which is in agreement with the asymptotic form of the exact Green's function with only Rashba spin-orbit coupling (see Ref.~\onlinecite{Imamura2004}), except that $q_F=\sqrt{k_F^2+m^2\alpha^2}$ is replaced here by $k_F$. This difference arises because in the derivation we neglected corrections of order $\alpha_\pm^2$, which implies that our result is only accurate for $R \ll k_F \lambda_\pm^2$. However, if the condition $k_F \lambda_\pm \gg 1$ is valid (which is often satisfied in practice), Eq.~(\ref{Gclean_largeR_RD}) becomes inaccurate only on a length scale much larger than $\min\{ \lambda_+,\lambda_-\}$.

By using Eqs.~(\ref{chi_leading}) and (\ref{Gclean_largeR_RD}), the following result for the susceptibility tensor is obtained at $T \to 0$
\begin{equation}\label{chi0_average_RD}
\overline{\chi_{ij}}({\bf R})=\frac{m\sin(2k_F R)}{2\pi^2 R^2} \, e^{-R/\ell} \, \mathcal{R}^{RD}_{ij}({\bf R}),
\end{equation}
where$ \mathcal{R}^{RD}_{ij}=\frac12 {\rm Tr}[\sigma_i U \sigma_j U^\dag]$ and
$U$ appears in the second line of Eq.~(\ref{Gclean_largeR_RD})  
\begin{equation}\label{U_def}
U=e^{i R m \boldsymbol{\alpha}_\varphi \cdot \boldsymbol{\sigma} }.
\end{equation}
Clearly, $U$ is a rotation operator of angle $2m\alpha_\varphi R$ around $\boldsymbol{\alpha}_\varphi$. This gives that $\mathcal{R}^{RD}$ is a rotation matrix, which generalizes the twisted exchange interaction of Ref.~\onlinecite{Imamura2004}. 

The effect of the impurity potential in Eq.~(\ref{chi0_average_RD}) is the factor $e^{-R/\ell}$. Since the disorder-averaged calculation is valid in the diffusive limit $R \gg \ell$, the final result for $\overline{\chi_{ij}}$ is exponentially small and it is necessary to consider the second-order correlations $\overline{\chi_{ij}\chi_{i'j'}}$ (see following sections). However, Eq.~(\ref{chi0_average_RD}) is still useful in the ballistic case, when the expression for $\chi_{ij}({\bf R})$ is formally identical to Eq.~(\ref{chi_leading}) if the $\tau \to \infty$ limit is taken in the Green's functions. Therefore, Eq.~(\ref{chi0_average_RD}) is also valid for $R\ll \ell$. In this case, both disorder-averaging and the exponential suppression can be dropped.

Finally, we note that the averaged charge response $\chi_{00}({\bf R})$ can also be simply calculated from Eq.~({\ref{Gclean_largeR_RD}) and is the same as in the absence of spin-orbit coupling. This is at variance with recent results of Ref.~\onlinecite{Badalyan2009}, where anisotropic Friedel oscillations  were obtained, which can also display beatings as function of $R$. We attribute this discrepancy to our approximation of small spin-orbit coupling. The effects of Ref.~\onlinecite{Badalyan2009} are quadratic in $\alpha_\pm$  at small spin-orbit couplings and thus are only relevant for distances much larger than $\lambda_\pm$, when Eq.~({\ref{Gclean_largeR_RD}) becomes inaccurate.

\section{Special limits of the second-order correlations}\label{sec_limits_correlations}

The general structure of $\overline{\chi_{ij}\chi_{i'j'}}$ following from Eq.~(\ref{correlations_formula}) can be represented as
\begin{equation}\label{chichi_compact}
\overline{\chi_{ij}\chi_{i'j'}}({\bf R}) = \frac{2 m^2}{3\pi^4 R^4} \, \eta_{ij,i'j'},
\end{equation}
where $\eta_{ij,i'j'}$ is defined in Eq.~(\ref{eta_generaldef}). We find that such a tensor is of order unity, which implies that the RKKY interaction decays in magnitude as $1/R^2$. 

As discussed in Appendix~\ref{app_delta13}, the only dependence of $\eta_{ij,i'j'}$ is from the ratios $R/\lambda_\pm$ and the direction of ${\bf R}$. This is similar to the results of Refs.~\onlinecite{Zyuzin1986,Jagannathan1988}, where the ratio $R/L_{\rm s.o.}$ appears instead of $R/\lambda_\pm$ ($L_{\rm s.o.}$ is the spin-orbit diffusion length, in the presence of spin-orbit scattering). While it is possible to evaluate explicitly the tensor $ \eta_{ij,i'j'}$ as function of $R/\lambda_\pm$ (see Sec.~\ref{sec_rashba_only} for pure Rashba spin-orbit coupling), we start discussing Eq.~(\ref{chichi_compact}) in some limiting cases that can be solved analytically and have a direct physical interpretation.

\subsection{Small spin-orbit coupling: $\ell \ll R \ll \lambda_\pm$}\label{sec_smallR}

In this case, since the relevant momenta and frequencies of $X_D({\bf q},\omega)$ are of order $q \sim 1/R$ and $\omega\tau \sim (\ell/R)^2 $, we can neglect the second line of Eq.~(\ref{X_D_result}) (proportional to the spin-orbit couplings) and the diffuson (\ref{diffuson_def}) can be evaluated analytically as\cite{Jagannathan1988}
\begin{equation}
D^{\mu\nu}_{\bf R}(w)=\frac{K_0(\frac{R}{\ell}\sqrt{2\tau\omega}  )}{2\pi m \tau \ell^2} \delta_{\mu\nu},
\end{equation}
where $K_0(z)$ is a modified Bessel function of the second kind.\cite{Abramowitz1972} Finally, Eq. (\ref{correlations_formula}) gives
\begin{equation}\label{corr_chi_noSO}
\overline{\chi_{ij}\chi_{i'j'}}({\bf R}) = \frac{2 m^2}{3\pi^4 R^4} \, \delta_{ij}\delta_{i'j'}.
\end{equation}
The variance of the RKKY energy immediately follows from $\overline{\chi_{ij}\chi_{i'j'}}$ [see Eq.~(\ref{Himp_definition})]
\begin{equation}\label{H2_noSO}
\overline{(H_{12})^2} =  \frac{2 J^4 m^2}{3\pi^4 R^4} ({\bf I}_1 \cdot {\bf I}_2)^2~.
\end{equation}
Therefore, the interaction between the magnetic moments is of the Heisenberg type. This is appropriate for a system without spin-orbit coupling\cite{Jagannathan1988} where the nonlocal polarization created by the impurity at the origin is everywhere parallel to ${\bf I}_1$. Equation~(\ref{H2_noSO}) reveals that such polarization is long-ranged and decays in magnitude as $1/R^2$. However, its strength is randomized in sign and the average susceptibility is exponentially suppressed, see Eq.~(\ref{chi0_average_RD}).

\subsection{Large spin-orbit coupling: $\ell  \ll \lambda_\pm \ll R$}\label{sec_bigR}

In the opposite limit of a large distance between the magnetic impurities (relative to the spin-orbit precession lengths $\lambda_{\pm}$), the spin-orbit contribution to Eq.~(\ref{X_D_result}) is much larger than the first line. Therefore, $[1-X_D]_{00}$ is much smaller than all the other matrix elements. Since the diffuson is obtained from $(1-X_D)^{-1}$ we can approximate Eq.~(\ref{X_D_result}) as
\begin{equation}
D^{\mu\nu}_{\bf R}(w)=\frac{K_0(\frac{R}{\ell}\sqrt{2\tau\omega}  )}{2\pi m \tau \ell^2} \,\delta_{\mu0}\delta_{\nu0},
\end{equation}
which gives
\begin{equation}\label{chi2_beta0}
\overline{\chi_{ij}\chi_{i'j'}}({\bf R}) = \frac{ m^2}{6\pi^4 R^4} \, \delta_{ii'}\delta_{jj'}.
\end{equation}
The variance of the RKKY energy reads in this case
\begin{equation}\label{H2_largeR}
\overline{(H_{12})^2}= \frac{J^4  m^2}{6\pi^4 R^4} {\bf I}_1^2  {\bf  I}_2^2~.
\end{equation}
Therefore, the magnetic interaction decays with the same power law $1/R^2$ as in the absence of spin-orbit couplings, see Eq.~(\ref{H2_noSO}). However, the variance becomes independent of the orientations of the two magnetic moments since ${\bf I}_{i}^2$ are constants. This fact becomes clear taking into account the spin relaxation of the electron spins induced by the spin-orbit interaction, which occurs on a length scale $\lambda_\pm$. The direction of the electron polarization induced by the first impurity is parallel to ${\bf I}_1$ close to its location (i.e., at the origin) but is fully randomized at a distance $R \gg \lambda_\pm$. Since the spin polarization induced in ${\bf R}$ by the first impurity is finite but distributed on average with spherical symmetry, Eq.~(\ref{H2_largeR}) can be simply understood. Notice also that Eq.~(\ref{H2_largeR}) is reduced by a factor 1/4 with respect to the maximum variance of Eq.~(\ref{H2_noSO}).

\subsection{The case $|\alpha|=|\beta|$}\label{sec_equalRD}

We consider here Rashba and Dresselhaus spin-orbit couplings with equal strength, when a particular component of the spin is conserved.\cite{Schliemann2003} For definiteness, we assume $\alpha=\beta$ (i.e., $\alpha_-=0$) such that $\sigma_y$ is conserved, while a similar argument is valid if $\alpha_+=0$. This situation is most simply treated using an exact unitary transformation since the electron Hamiltonian (\ref{so_hamilton}) can be expressed as
\begin{equation}
H_{el}=  U^\dag \left( \frac{{\bf p}^2}{2m}+V({\bf r})\right) U + \frac{m\alpha_+^2}{2},
\end{equation}
independently of the disorder potential $V({\bf r})$, where $U=e^{-i (m\alpha_+) x\sigma_y }$ is the operator of Eq.~(\ref{U_def}). Therefore, for a given disorder potential, $\chi_{ij}({\bf 0},{\bf R})$ is obtained from the result without spin-orbit coupling after a rotation of ${\bf I}_2$ (and a small shift of the chemical potential $\epsilon_F$). Such rotation is around ${\bf e}_y$ by an angle $2 m\alpha_+ R\cos\varphi$.

Since this exact property holds for every disorder realization, it still holds after averaging. The result for the susceptibility tensor is in agreement with Eq.~(\ref{chi0_average_RD}) while the second-order correlations are
\begin{equation}\label{chi2_betaalpha}
\overline{\chi_{ij}\chi_{i'j'}}({\bf R}) = \frac{2 m^2}{3\pi^4 R^4} \, \mathcal{R}^{RD}_{ij}\mathcal{R}^{RD}_{i'j'},
\end{equation}
with $ \mathcal{R}^{RD}_{ij}$ the same rotation matrix of Eq.~(\ref{chi0_average_RD}), evaluated at $\alpha_-=0$. The variance of the energy is
\begin{equation}
\overline{(H_{12})^2} =  \frac{2 J^4 m^2}{3\pi^4 R^4} ({\bf I}_1 \cdot \mathcal{R}^{RD}{\bf I}_2)^2~.
\end{equation}
The same result can be obtained by direct evaluation of Eq.~(\ref{correlations_formula}), see Appendix~\ref{app_ReqD}.

If the spin orbit couplings are not exactly equal, Eq.~(\ref{chi2_betaalpha}) is only valid for $R \ll \lambda_-$ and, by increasing $R$, a smooth crossover to Eq.~(\ref{chi2_beta0}) is realized when $R \gg \lambda_-$. Notice however that our treatment is only valid if $R$ is within the phase-coherence length of the electron gas. 

The regime close to $\alpha=\beta$ was also examined in Refs.~\onlinecite{Chalaev2009} and \onlinecite{Duckheim2010}, where it was found that several linear response functions have a non-analytic behavior. For example, the electrically induced spin-polarization of a phase coherent sample of size $L$ is zero or finite depending on the order of the limits $L \to \infty$ and $\alpha \to \beta$.\cite{Duckheim2010} Similarly, we find in our case that $\overline{\chi_{ij}\chi_{i'j'}}$ has two different forms, Eqs.~(\ref{chi2_beta0}) and (\ref{chi2_betaalpha}), depending on the order of the $R\to \infty$ and $\lambda_- \to \infty$ limits. 

\section{Variance of the RKKY interaction with only Rashba coupling}\label{sec_rashba_only}

In the general case, we could not obtain analytic results for Eq.~(\ref{correlations_formula}). We then resort to direct numerical integration and consider here the specific case of pure Rashba spin-orbit coupling ($\beta=0$). Since the Rashba spin-orbit coupling has rotational symmetry in the $xy$-plane, we assume in the following ${\bf R} = R {\bf e}_x$. Furthermore, results for pure Dresselhaus spin-orbit coupling can be simply obtained applying a spin rotation. The general form of the correlation tensor $\overline{\chi_{ij}\chi_{i'j'}}$ is given in Eq.~(\ref{chichi_compact}) in terms of the tensor $\eta_{ij,i'j'}$, which uniquely depends on the ratio $R/\lambda$ (since for Rashba spin-orbit coupling $\lambda_\pm =\lambda$). Details on the evaluation of the functions $\eta_{ij,i'j'}$ are given in Appendix~\ref{app_delta13} while we find it more transparent to discuss here the variance of the RKKY interaction energy instead of the full correlation tensor. We first discuss specific contributions to the variance, corresponding to definite orientations of ${\bf I}_{1,2}$ in the classical limit. A generally applicable expression Eq.~(\ref{general_final}) is obtained at the end of this section.

\subsection{${\bf I}_1$ and ${\bf I}_2$ parallel to ${\bf e}_y$}

As a first example, we consider the contribution to the variance involving only $I_{1y}$ and $I_{2y}$, i.e., due to the $\overline{\chi_{yy}\chi_{yy}}$ component of the correlations tensor:
\begin{equation}\label{deltaH_22}
\left[ \overline{(H_{12})^2}\right]_{y,y}=\frac{2J^4 m^2}{3\pi^4 R^4}   A_0 I_{1y}^2 I_{2y}^2.
\end{equation}
From Eq.~(\ref{chichi_compact}), it is clear that we defined $A_0=\eta_{yy,yy}$, which is plotted in Fig.~\ref{M0} and is a monotonically decreasing function of $R/\lambda$, from 1 to 1/4. Therefore, it correctly interpolates between the two limits of Secs. \ref{sec_smallR} and \ref{sec_bigR}. We note that the asymptotic value 1/4 is reached with good approximation already at moderate values of $R/\lambda \gtrsim 2$. The reason is that the asymptotic value is approached exponentially, as it can be seen in the inset of Fig.~\ref{M0}}. A similar behavior was found for spin-orbit scattering, where the exponentially small corrections in the limit $R/L_{\rm s.o.} \gg 1$ were explicitly evaluated.\cite{Zyuzin1986,Jagannathan1988}

\begin{figure}
\begin{center}
\includegraphics[width=0.4\textwidth]{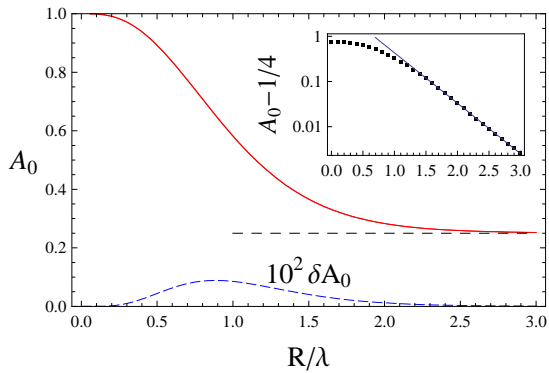}
\caption{\label{M0} (Color online.) Main plot: dependence of $A_0$ of Eq.~(\ref{deltaH_22}) on $R/\lambda$ (red solid curve). The horizontal dashed line is the asymptotic value $1/4$. The coefficient $A'_0$ of Eq.~(\ref{xzxz_final}) almost perfectly coincides with $A_0$. The blue dashed curve is the small difference $\delta A_0=A_0-A_0'$ (multiplied by a factor of 100). The inset shows $A_0-1/4$ (dots) on a semilogarithmic plot. A linear fit gives $\ln(A_0-1/4)\simeq 1.73 - 2.57 (R/\lambda)$.}
\end{center}
\end{figure}

\subsection{${\bf I}_1$ along ${\bf e}_y$ and ${\bf I}_2$ in the perpendicular plane}\label{sec_I12_perp}

As a second case, we consider terms involving only $I_{1y}$ and the $x,z$ components of ${\bf I}_2$:
\begin{eqnarray}\label{H12_yxz}
\left[ \overline{(H_{12})^2}\right]_{y,xz}= \frac{2J^4 m^2}{3\pi^4 R^4}  I_{1y}^2 
(\eta_{yx,yx} I_{2x}^2+  \eta_{yz,yz}   I_{2z}^2  \nonumber \\
 + \eta_{yx,yz} [ I_{2x}, I_{2z}]_+ ),
 \end{eqnarray}
where $[a,b ]_\pm=ab \pm ba$. We do not provide here explicit expressions for the components of $\eta_{ij,i'j'}$ entering Eq.~(\ref{H12_yxz}) but they can be found as discussed in Appendix~\ref{app_delta13}. In particular, they follow directly from Eq.~(\ref{eta_generaldef}). Equation (\ref{H12_yxz}) can be directly applied to classical magnetic moments if ${\bf I}_1$ along ${\bf e}_y$ and ${\bf I}_2$ in the $xz$-plane, and we can simply set $[ I_{2x}, I_{2z}]_+=2I_{2x} I_{2z}$ in this case. More generally, this is not justified since ${\bf I}_{1,2}$ are angular momentum operators and Eq.~(\ref{H12_yxz}) represents a specific contribution to the total variance Eq.~(\ref{general_final}). 

Eq.~(\ref{H12_yxz}) can be simplified with a rotation of ${\bf I}_2$ along ${\bf e}_y$ by defining
\setlength\arraycolsep{2pt}
\begin{equation}\label{S_def}
\left(
\begin{array}{c}
S_{2x} \\
S_{2z}
\end{array}
\right)=
\left(
\begin{array}{cc}
\cos\Phi & -\sin\Phi\\
\sin\Phi & \cos\Phi
\end{array}
\right)
\left(
\begin{array}{c}
I_{2x} \\
I_{2z}
\end{array}
\right),
\end{equation}
where $\Phi$ satisfies the condition
\begin{equation}\label{Phi_def}
\tan 2\Phi =\frac{2\eta_{yx,yz}}{\eta_{yz,yz}- \eta_{yx,yx}}.
\end{equation}
The plot of $\Phi$ as function of $R/\lambda$ is given in Fig.~\ref{figure_Phi}. After this transformation, the anticommutator term drops out and Eq.~(\ref{H12_yxz}) becomes
\begin{eqnarray}\label{yxz_final}
\left[ \overline{(H_{12})^2}\right]_{y,xz}= \frac{2J^4 m^2}{3\pi^4 R^4}  (A_z  I_{1y}^2 S_{2x}^2+A_x  I_{1y}^2 S_{2z}^2),
\end{eqnarray}
 where the coefficients $A_x$ and $A_z$ smoothly interpolate from $0$ to $1/4$ and are plotted in Fig.~\ref{figure_Mi}.

\begin{figure}
\begin{center}
\includegraphics[width=0.4\textwidth]{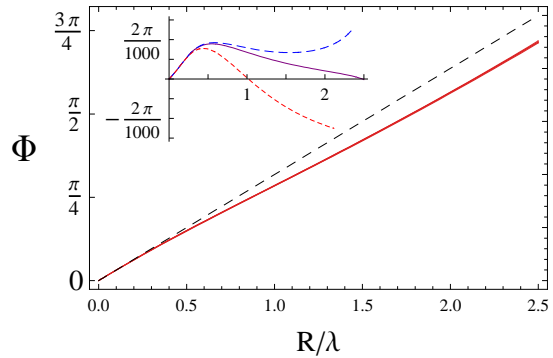}
\caption{\label{figure_Phi}(Color online.) Main plot: the four solid red overlapping curves show the dependence on $R/\lambda$ of the rotation angles $\Phi$, $\Phi'$, and $\Phi^\pm$ defined in Eqs.~(\ref{Phi_def}), (\ref{Phi1_def}), and (\ref{Phi_pm_def}), respectively. The dashed line indicates the linear dependence $\Phi=R/\lambda = m\alpha R$. Inset: the small differences $\Phi'-\Phi$ (purple, solid), $\Phi^+-\Phi$ (blue, long dashed), and $\Phi^--\Phi$ (red, short dashed) are plotted as functions of $R/\lambda$.  }
\end{center}
\end{figure}

%
\begin{figure}
\begin{center}
\includegraphics[width=0.4\textwidth]{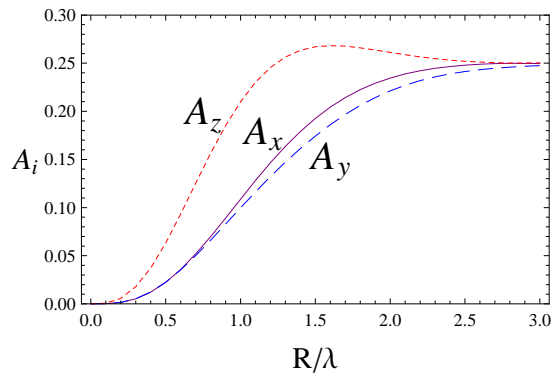}
\caption{\label{figure_Mi} (Color online.) Dependence on $R/\lambda$ of $A_x$ (purple, solid), $A_y$ (blue, long dashed), and $A_z$ (red, short dashed). }
\end{center}
\end{figure}

If the magnetic moments ${\bf I}_1$ and ${\bf I}_2$ are interchanged, we can proceed in a similar way. From Appendix \ref{eta_simmetries} we have that $\eta_{xy,zy}=-\eta_{yx,yz}$, implying that the final expression is simplified by rotating ${\bf I}_1$ by an angle $-\Phi$. This gives
\begin{eqnarray}\label{xzy_final}
\left[ \overline{(H_{12})^2}\right]_{xz,y}= \frac{2J^4 m^2}{3\pi^4 R^4}   (A_x S_{1z}^2 I_{2y}^2+A_z S_{1x}^2 I_{2y}^2)  .
 \end{eqnarray}
Since ${\bf I}_1$ and ${\bf I}_2$ are transformed with opposite angles, the two rotated reference frames differ by $2\Phi$. From Fig.~\ref{figure_Phi} we see that $2\Phi \simeq 2 m\alpha R$, which is reminiscent of the twisted exchange interaction of Ref.~\onlinecite{Imamura2004} and Eq.~(\ref{chi0_average_RD}). 

\subsection{${\bf I}_1$ and ${\bf I}_2$ perpendicular to ${\bf e}_y$}\label{sec_I12_both_perp}

Consider now the contributions from $\overline{\chi_{ij}\chi_{i'j'}}$ if all the indexes are different from $y$. As it turns out, not all such components are independent. Besides the symmetries discussed in Appendix \ref{eta_simmetries} we find $\eta_{xx,xx}=\eta_{zz,zz}$ and $\eta_{xx,xz}=\eta_{zz,xz}$. The three other relevant components of the tensor are $\eta_{xz,xz}$, $\eta_{xz,zx}$, and $\eta_{xx,zz}$. Therefore, the following expression is obtained
\begin{eqnarray}\label{xzxz_variance}
\left[ \overline{(H_{12})^2}\right]_{xz,xz}&&   =\frac{J^4 m^2}{3\pi^4 R^4}\Big\{V_1^T \mathcal{M} V_2 \nonumber \\
+ ( \eta_{xx,xx} &&  + \eta_{xz,xz}) (I_{1x}^2 + I_{1z}^2) (I_{2x}^2 + I_{2z}^2) \nonumber \\
+ (   \eta_{xx,zz} && - \eta_{xz,zx})[I_{1x},I_{1z}]_- [I_{2x},I_{2z}]_- \Big \},
\end{eqnarray}
where we defined 
\begin{eqnarray}
&&
V_i =
\left(
\begin{array}{c}
I_{ix}^2-I_{iz}^2 \\
{[} I_{ix}, I_{iz}]_+
\end{array}
\right), \\
\nonumber \\
&& \mathcal{M} =\left( 
\begin{array}{cc}
\eta_{xx,xx}-\eta_{xz,xz} & 2 \eta_{xx,xz} \\
-2\eta_{xx,xz}  & \eta_{xx,zz}+\eta_{xz,zx}
\end{array}
\right),
\end{eqnarray}
and $V_1^T$ is the transpose of $V_1$. By a suitable rotation of ${\bf I}_2$ as in Eq.~(\ref{S_def}) and of ${\bf I}_1$ by an opposite angle, the expression $V_1^T \mathcal{M} V_2$ can be diagonalized. However, the angle is not exactly $\Phi$ as before, but a slightly different value $\Phi^\prime$ which satisfies
\begin{equation}\label{Phi1_def}
\tan 4\Phi' =\frac{4 \eta_{xx,xz}}{\eta_{xz,xz}-\eta_{xx,xx} - \eta_{xx,zz}-\eta_{xz,zx}}.
\end{equation}
As seen in Fig.~\ref{figure_Phi}, the plot of $\Phi$ and $\Phi'$ are practically indistinguishable. The dependence of the small difference $\delta\Phi'=\Phi'-\Phi$ on $R/\lambda$ is shown in the inset of Fig.~\ref{figure_Phi}. In terms of the rotated magnetic moments ${\bf S}_i^\prime \simeq {\bf S}_i$, Eq.~(\ref{xzxz_variance}) takes the form
\begin{eqnarray}\label{xzxz_final}
\left[ \overline{(H_{12})^2}\right]_{xz,xz}= \frac{2J^4 m^2}{3\pi^4 R^4} && \Big\{ \sum_\pm  B_y^\pm[ S_{1x}^{\prime}, S_{1z}^{\prime}]_\pm[ S_{2x}^{\prime}, S_{2z}^{\prime}]_\pm \nonumber \\
&&+A'_0  (S_{1x}^{\prime 2}S_{2x}^{\prime 2}+S_{1z}^{\prime 2}S_{2z}^{\prime 2}) \nonumber \\
&&+A_y (S^{\prime 2}_{1x} S^{\prime 2}_{2z}+S^{\prime 2}_{1z} S^{\prime 2}_{2x})  \Big\}.
\end{eqnarray}
The coefficients $A_y$, $B_y^+$, and $B_y^-$ are plotted in Figs. \ref{figure_Mi}, \ref{figure_Bip}, and \ref{figure_Bim}, respectively. Finally, the value of $A_0'$ is very similar to $A_0$: its plot and the small difference $\delta A_0=A_0-A_0'$ are shown Fig.~\ref{M0}. As discussed in more detail in Appendix~\ref{app_delta13}, the fact that the angles $\Phi$ and $\Phi'$ and the amplitudes $A_0$ and $A_0'$ are very similar can be ascribed to the same reason, i.e., the diffuson matrix $D_{\bf R}(\omega)$ can be diagonalized at each value of $R/\lambda$ with a rotation which is almost independent of the frequency $\omega$. The rotation angle is approximately equal to $\Phi$.

\begin{figure}
\begin{center}
\includegraphics[width=0.4\textwidth]{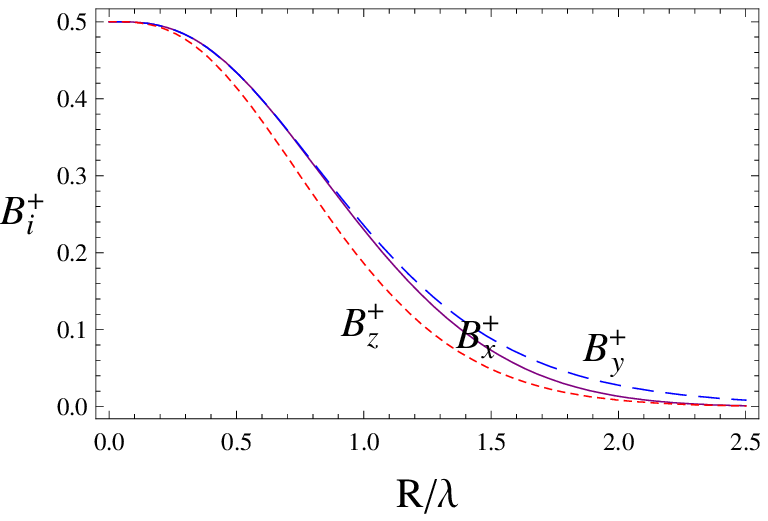}
\caption{\label{figure_Bip} (Color online.) Dependence  on $R/\lambda$ of $B_x^+$ (purple, solid), $B_y^+$ (blue, long dashed), and $B_z^+$ (red, short dashed). }
\end{center}
\end{figure}

\begin{figure}
\begin{center}
\includegraphics[width=0.4\textwidth]{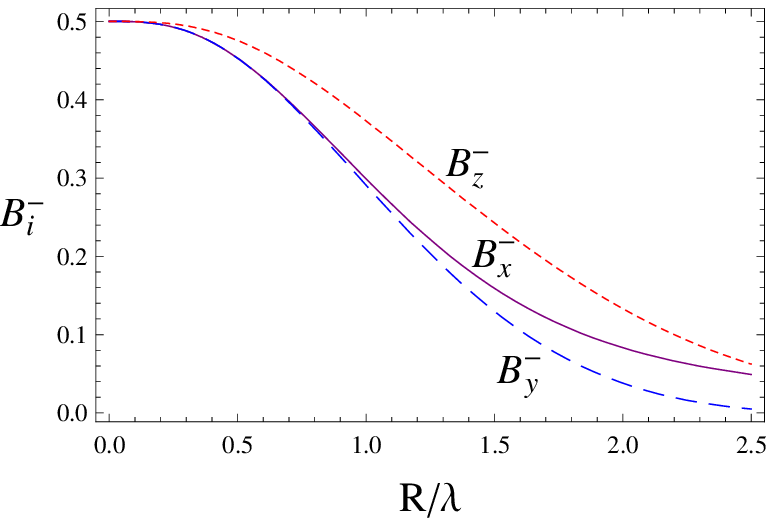}
\caption{\label{figure_Bim} (Color online.) Dependence  on $R/\lambda$ of $B_x^-$ (purple, solid), $B_y^-$ (blue, long dashed), and $B_z^-$ (red, short dashed). }
\end{center}
\end{figure}

\subsection{General case}

We consider now an expression valid for generic orientations of the magnetic moments. Such generic formula for the variance contains the sum of Eqs.~(\ref{deltaH_22}), (\ref{yxz_final}), (\ref{xzy_final}), and (\ref{xzxz_final}), but an additional term $\Delta  \overline{(H_{12})^2}$ appears, involving the remaining six independent elements of the correlation tensor $\eta_{ij,i'j'}$. This last contribution is expressed in compact form as
\begin{equation}\label{interference}
\Delta  \overline{(H_{12})^2}=\frac{J^4 m^2}{3\pi^4 R^4}\sum_\pm V_{1\pm}^{T} \mathcal{M}_\pm V_{2\pm} ,
\end{equation}
where we defined 
\setlength\arraycolsep{5pt}
\begin{eqnarray}
&&
V_{i\pm} =
\left(
\begin{array}{c}
{[} I_{iy}, I_{ix}]_\pm \\
{[} I_{iy}, I_{iz}]_\pm
\end{array}
\right), \\
\nonumber \\
&& \mathcal{M}_\pm =\left( 
\begin{array}{cc}
\eta_{xx,yy} \pm \eta_{xy,yx} & \eta_{xz,yy} \pm \eta_{xy,yz} \\
-\eta_{xz,yy} \mp \eta_{xy,yz}  & \eta_{zz,yy} \pm \eta_{zy,yz}
\end{array}
\right).
\end{eqnarray}
As before, the matrices $\mathcal{M}_\pm$ can be diagonalized with opposite rotations of ${\bf I}_1$ and ${\bf I}_2$. The angles are different for the two terms entering Eq.~(\ref{interference}). They are given by
\begin{equation}\label{Phi_pm_def}
\tan2\Phi^\pm =\frac{-2(\eta_{xz,yy} \pm \eta_{xy,yz})}{\eta_{xx,yy} \pm \eta_{xy,yx} + \eta_{zz,yy} \pm \eta_{zy,yz}},
\end{equation}
and are plotted in Fig.~\ref{figure_Phi}. As seen, $\Phi^\pm$ are both almost identical to $\Phi$ of Eq.~(\ref{Phi_def}). The differences $\delta\Phi^\pm=\Phi^\pm-\Phi$ are plotted in the inset of Fig.~\ref{figure_Phi}, which shows that $\delta\Phi^\pm$ are in general very small and only increase in magnitude at large values of $R/\lambda$. However, as discussed in Sec.~\ref{sec_bigR}, the angular dependence of the magnetic moments becomes unimportant in this limit. In terms of the rotated magnetic moments ${\bf S}_i^\pm \simeq {\bf S}_i$, we can rewrite Eq.~(\ref{interference}) as follows
\begin{eqnarray}\label{interferenc_final}
\Delta  \overline{(H_{12})^2}=\frac{2J^4 m^2}{3\pi^4 R^4}   \sum_\pm   \Big\{ B_x^\pm[ S_{1y}^{\pm}, S_{1z}^{\pm}]_\pm[ S_{2y}^{\pm}, S_{2z}^{\pm}]_\pm \nonumber \\
+B_z^\pm[ S_{1x}^{\pm}, S_{1y}^{\pm}]_\pm[ S_{2x}^{\pm}, S_{2y}^{\pm}]_\pm \Big\}, \qquad
\end{eqnarray}
where the coefficients $B_{x,z}^+$, and $B_{x,z}^-$ are plotted in Figs. \ref{figure_Bip} and \ref{figure_Bim}, respectively.

\begin{figure}
\begin{center}
\includegraphics[width=0.4\textwidth]{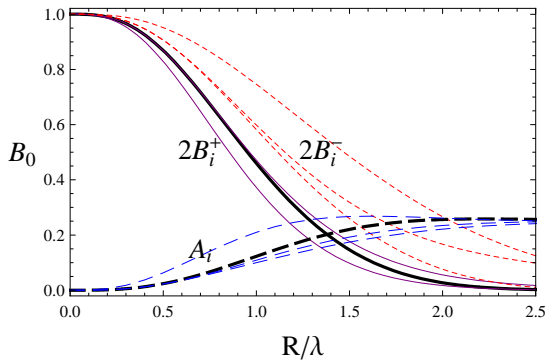}
\caption{\label{figure_B0} (Color online.) Thick black solid line: plot of $B_0=2 B_x^+$ as function of $R/\lambda$. This can be compared to the coefficients $2B_i^+$ (purple solid lines, see also Fig.~\ref{figure_Bip}) and $2B_i^-$  (short dashed red lines, see also Fig.~\ref{figure_Bim}). Thick black dashed line:  plot of $A_0-B_0$, which can be compared to the $A_i$ coefficients (long dashed blue lines, see also Fig.~\ref{figure_Mi}).}
\end{center}
\end{figure}

Finally, we can summarize the final result for the variance as follows
\begin{eqnarray}\label{general_final}
\overline{(H_{12})^2} \simeq  &&  \frac{2J^4 m^2}{3\pi^4 R^4}  \Big\{ A_0 S_{1x}^2S_{2x}^2+ A_x(S_{1y}^2S_{2z}^2+S_{1z}^2S_{2y}^2) \nonumber \\
&&  +\sum_\pm B_x^\pm  [ S_{1y}, S_{1z}]_\pm[ S_{2y}, S_{2z}]_\pm   + {\rm c.p.}  \Big\},
\end{eqnarray}
where c.p. indicates the remaining terms, obtained by cyclic permutations of the indexes $xyz$. Two such terms are of the form $A_0 S_{1i}^2 S_{2i}^2$ ($i=y,z$) and the other six correspond to the coefficients $A_{y}, A_{z}$ and $B_{y}^\pm,B_{z}^\pm$. Notice that in Eq.~(\ref{general_final}) we neglected the small difference between $A_0$ and $A_0'$ as well as between ${\bf S}_i$ and ${\bf S}_i^\prime$, ${\bf S}_i^\pm$. If we do not use this approximation and we apply Eq.~(\ref{general_final}) to spin-1/2 magnetic impurities we obtain
\begin{equation}\label{general_final_12}
\overline{(H_{12})^2} \simeq   \frac{J^4 m^2}{3\pi^4 R^4}\bigg[ \frac{A_0}{2} +A_0^\prime + \sum_{i=x,y,z} \left(   A_i - 2B_i^-  S^-_{1i} S^-_{2i} \right) \bigg],
\end{equation}
by using $S_{iy}^\prime = S_{iy}^- = I_{iy}$.

For the opposite case of classical variables, the commutator terms in Eq.~(\ref{general_final}) vanish. To further simplify the expression, one can set $B^+_{x,y,z}\simeq B_x^+ \equiv B_0/2$ and $A_{x,y,z}\simeq A_0-B_0$. The error introduced by these approximations is seen in Fig.~\ref{figure_B0}, where we plot $B_0$ and $A_0-B_0$ together with the corresponding coefficients $2B_i^+$ and $A_i$ of the previous Figs.~\ref{figure_Bip} and \ref{figure_Mi}. Equation~(\ref{general_final}) then assumes the following simple form
\begin{equation}\label{general_final_classic}
\overline{(H_{12})^2} \simeq   \frac{2J^4 m^2}{3\pi^4 R^4}   \left[  B_0 ({\bf S}_{1}\cdot {\bf S}_2)^2  + (A_0-B_0) {\bf S}_1^2 {\bf S}_2^2 \right],
\end{equation}
which explicitly interpolates between Eq.~(\ref{H2_noSO}) at $R/\lambda \ll 1$ (when $A_0=B_0=1$) and Eq.~(\ref{H2_largeR}) at $R/\lambda \gg 1$ (when $A_0=1/4$ and $B_0=0$). Equation~(\ref{general_final_classic}) can also be used to approximate the quantum case, if both the commutator and anticommutator coefficients are set equal to $B_0/2$. This can be justified from the fact that all the $2B_i^\pm$ coefficients decrease from $1$ to $0$ on a similar scale of  $R/\lambda$. However, the approximation $B_i^-\simeq B_0/2$ is significantly worse than for $B_i^+$ (see Fig.~\ref{figure_B0}).

\section{Conclusion}\label{the_end}

We discussed in this paper the influence of non-magnetic disorder and spin-orbit couplings on the RKKY interaction in two dimensions. Our analysis focused on the effect of the Rashba and Dresselhaus spin-orbit interactions, rather than spin-orbit scattering.\cite{Jagannathan1988} We examined both the ballistic and disorder-averaged susceptibility. Furthermore, we characterized the second-order correlations of the susceptibility tensor.

For the susceptibility tensor, we extended the result found in Ref.~\onlinecite{Imamura2004} and obtained that the RKKY interaction has the form of a twisted Heisenberg exchange for generic couplings $\alpha,\beta$. An interesting question is if such form of twisted exchange coupling is still valid at large spin-orbit couplings or distances, beyond the order of approximation considered here. In this regime, anisotropic Friedel oscillations and the presence of beatings were obtained in the density response.\cite{Badalyan2009} 

At distance larger than the mean-free-path, the disorder-averaged RKKY interaction is exponentially suppressed. Therefore, it has to be characterized through the second-order correlations of the susceptibility tensor, which decay with the same power law $1/R^4$ of the clean case. We obtained the tensor structure explicitly in a number of interesting limits and presented a detailed study with pure Rashba spin-orbit coupling. In this case, the final result Eq.~(\ref{general_final}) is rather involved but the main features are well reproduced by a simpler approximate formula, Eq.~(\ref{general_final_classic}). We obtain that: (i) also for $R \gg \ell$ the RKKY interaction (in this case, its variance) can be expressed in terms of rotated magnetic moments. The relative angle deviates from the linear dependence of the clean case (see Fig.~\ref{figure_Phi}). (ii) The regime of large spin-orbit coupling is realized at moderate distance. As exemplified in Fig.~\ref{M0}, the coefficients of Eq.~(\ref{general_final}) approach their asymptotic values already at $R/\lambda \gtrsim 2 $ and the corrections are exponentially small.

We provide next a summary of relevant length scales and experimental conditions to which our results apply. First of all, it is required for all the distances involved in the problem to be smaller than the phase-coherence length $L_\varphi$ (which can be of the order of a few micrometers\cite{Ferrier2004}), the size $L$ of the sample, and the thermal diffusion length\cite{Bulaevskii1986}  ($L_T=\sqrt{D/\pi T}$, with $D=v_F^2\tau/2$ the diffusion constant). We also generally assumed throughout the paper that $1/k_F$ is the smallest length.
For magnetic semiconductors this implies that, since $R\gg 1/k_F$,  the free carriers should be provided mainly form external doping and not by the magnetic ions. The impurity calculations are generally performed in the diffusive regime, which implies $R \gg \ell$. Finally, effects of the spin-orbit coupling are relevant when the spin-orbit lengths become comparable to the distance $R$. Therefore, we can summarize all these conditions as follows:
\begin{equation}
\min\{L_\varphi, L, L_T \} \gg R, \lambda_\pm \gg \ell \gg 1/k_F. 
\end{equation}
Additionally, our result for $\overline{\chi_{ij}}({\bf R})$ can be applied to the ballistic case. This is because the Green's functions appearing in the bubble diagram Eq.~(\ref{chi_leading}) become equal to the unperturbed ones in the limit $\tau \to \infty$. Therefore, Eq.~(\ref{chi0_average_RD}) becomes a valid expression for $\chi_{ij}({\bf R})$ (without disorder-averaging) if 
\begin{equation}
\min\{L_\varphi, L, L_T , \ell , \lambda_\pm^2 k_F \} \gg R, \lambda_\pm \gg 1/k_F. 
\end{equation}

In conclusion, our study is relevant to characterize magnetic structures and spin-glass properties in two-dimensional systems. However, instead of having several magnetic impurities, a single isolated magnetic moment could be considered. The spin polarization pattern induced in the electron gas could be revealed for surface states by scanning tunneling microscopy (STM) techniques.\cite{Wahl2007,Meier2010} STM studies also exist in magnetic semiconductors\cite{Richardella2010} but optical methods might represent a more viable option.\cite{Burch2008} In particular, spatially resolved optical imaging of the spin polarization is possible in semiconductor quantum wells.\cite{Sih2005} Optical pumping could also be exploited to polarize the central magnetic impurity or a small region of nuclear spins in a chosen direction. Finally, an alternative approach might be provided by spin-grating techniques,\cite{Weber2007} to obtain the spin susceptibility in momentum space.

\begin{acknowledgments}

We acknowledge useful discussions with O. Chalaev, M. Duckheim, D. L. Maslov, and M. J. Schmidt. This work was supported by
the Swiss NSF and the NCCR Nanoscience Basel.
\end{acknowledgments}

\appendix

\section{Cooperon and diffuson contributions to $\overline{\chi_{ij}\chi_{i'j'}}$}\label{appendix_cooperon_fluctuations}

\begin{figure}
\includegraphics[width=0.3\textwidth]{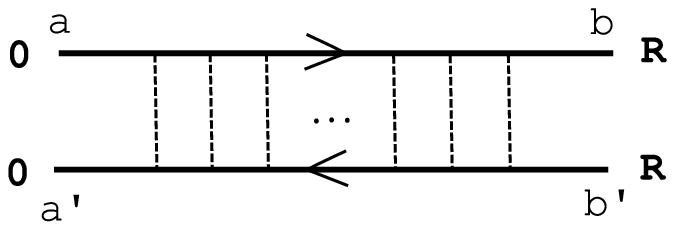}
\includegraphics[width=0.3\textwidth]{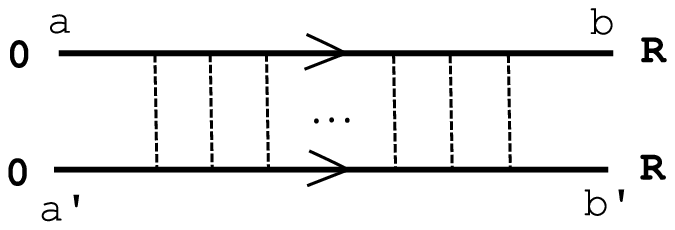}
\caption{\label{fig_cooper_ladder} Diffuson (top) and cooperon (bottom) ladder diagrams corresponding to Eqs.~(\ref{ladder_diff_def}) and (\ref{ladder_cooper_def}), respectively.}
\end{figure}

We discuss here in more detail the derivation of Eq.~(\ref{correlations_formula}) for $\overline{\chi_{ij}\chi_{i'j'}}$. The final result involves both diffuson and cooperon diagrams (see Fig.~\ref{diagrams_variance}), constructed from ladders schematically illustrated in Fig.~\ref{fig_cooper_ladder}. The diffuson ladder corresponds to the analytic expression
\begin{eqnarray}\label{ladder_diff_def}
\int  \frac{d{\bf q}}{(2\pi)^2}    \frac{e^{-i {\bf q}\cdot {\bf R}}}{(m\tau)^{n-1}}&& \int\prod_{i=1}^n  \frac{d{\bf k}_i }{(2\pi)^{2}} 
 \left[G_{\omega_1}({\bf k}_1) \ldots G_{\omega_1}({\bf k}_n) \right]_{ab}  \nonumber \\
\times && 
\left[G_{\omega_2}({\bf k}_{n}-{\bf q})\ldots G_{\omega_2}({\bf k}_{1}-{\bf q})  \right]_{b'a'} , \qquad
\end{eqnarray}
while for the cooperon ladder we have
\begin{eqnarray}\label{ladder_cooper_def}
\int  \frac{d{\bf q}}{(2\pi)^2}   \frac{e^{-i {\bf q}\cdot {\bf R}}}{(m\tau)^{n-1}}&& \int\prod_{i=1}^n  \frac{d{\bf k}_i }{(2\pi)^{2}} 
\left[G_{\omega_1}({\bf k}_1) \ldots G_{\omega_1}({\bf k}_n) \right]_{ab}  \nonumber \\
 \times && \left[G_{\omega_2}({\bf q}-{\bf k}_{1})\ldots G_{\omega_2}({\bf q}-{\bf k}_{n})  \right]_{a'b'} . \qquad
\end{eqnarray}
In Eqs.~(\ref{ladder_diff_def}) and (\ref{ladder_cooper_def}) the Green's functions in the square parentheses are $2\times 2$ matrices (in the spin space), $a,b,a',b'$ are spin indexes, and $(n-1)$ is the number of impurity lines ($n \geq 2$). We now make repeated use use of the identity\cite{Chalaev2009}
\begin{equation}
\delta_{ab}\delta_{cd}  =  \frac12 \sum_{\mu=0}^3 [\sigma_{\mu}]_{ac} [\sigma_{\mu}]_{db}, 
\end{equation}
for the diffuson ladder and
\begin{equation}
\delta_{ab}\delta_{cd}  =  \frac12 \sum_{\mu=0}^3 [\bar\sigma_{\mu}]_{ac} [\bar\sigma_{\mu}]_{db}, 
\end{equation}
for the cooperon, where $\bar\sigma_\mu = \sigma_y \sigma_\mu$. This allows to simplify (\ref{ladder_diff_def}) and (\ref{ladder_cooper_def}) as follows 
\begin{eqnarray}\label{ladder_diff}
&&\frac{m\tau}{2}\sum_{\mu\nu} [\sigma_{\mu}]_{aa'}[\sigma_{\nu}]_{b'b}\int \frac{d{\bf q}}{(2\pi)^2}  e^{-i {\bf q}\cdot {\bf R}} [(X_{D}({\bf q},\omega))^n] _{\mu\nu}, \qquad \\
\label{ladder_cooper}
&&\frac{m\tau}{2}\sum_{\mu\nu} [\bar\sigma^\dag_{\mu}]_{aa'}[\bar\sigma_{\nu}]_{b'b}\int \frac{d{\bf q}}{(2\pi)^2}  e^{-i {\bf q}\cdot {\bf R}} [(X_{C}({\bf q},\omega))^n] _{\mu\nu}, \qquad
\end{eqnarray}
where $X_{D}({\bf q},\omega)$ is defined in Eq.~(\ref{XD_def}) and 
\begin{equation}\label{XC_def}
X_{C}^{\mu\nu}({\bf q},\omega)=\frac{1}{2m\tau}\int \frac{d{\bf k}}{(2\pi)^2} {\rm Tr}[\bar\sigma_\mu G_{\omega_1}({\bf k}) \bar\sigma^\dag_\nu G^T_{\omega_2}({\bf q}-{\bf k})]. \nonumber
\end{equation}
In the above equations, $\omega=|\omega_1-\omega_2|$ and $G^T_\omega({\bf k})$ is the transposed Green's function. 

By summing Eqs.~(\ref{ladder_diff}) and (\ref{ladder_cooper}) over $n$ 
we obtain
$(m\tau)^2 [\sigma_{\mu}]_{aa'}[\sigma_{\nu}]_{b'b} D^{\mu\nu}_{\bf R}(\omega) $ for the diffuson and
$(m\tau)^2 [\bar\sigma^\dag_{\mu}]_{aa'}[\bar\sigma_{\nu}]_{b'b} C^{\mu\nu}_{\bf R}(\omega)$ for the cooperon case. The cooperon propagator $C_{\bf R}(\omega)$ is defined similarly to Eq.~(\ref{diffuson_def}), with $X_{C}$ instead of $X_D$. The full series of fluctuations diagram is than easily obtained
\begin{eqnarray}
\overline{\chi_{ij}\chi_{i'j'}}({\bf R})=   &&  \sum_{\mu, \nu, \mu', \nu'} \sum_{\omega_1,\omega_2}  (m\tau)^4 T^2  \nonumber \\
\times [ D^{\mu\nu}_{\bf R}(\omega)D^{\mu'\nu'}_{-\bf R}(\omega) && {\rm Tr}(\sigma_i \sigma_\mu \sigma_{i'}\sigma_{\nu'})
{\rm Tr}(\sigma_j \sigma_{\mu'}\sigma_{j'}\sigma_{\nu}) \qquad \nonumber \\
 + C^{\mu\nu}_{\bf R}(\omega)C^{\mu'\nu'}_{-\bf R}  (\omega) && {\rm Tr}(\sigma_i \bar\sigma^\dag_\mu \sigma^T_{i'}\bar\sigma_{\nu'}){\rm Tr}(\sigma_j \bar\sigma^\dag_{\mu'}\sigma^T_{j'}\bar\sigma_{\nu})]. \qquad 
\end{eqnarray}
We can finally take the limit of zero temperature. This amounts to replace $T^2\sum_{\omega_1,\omega_2}\to \frac{1}{2\pi^2}\int_0^\infty \omega d\omega$, using that the integrand depends on $\omega=|\omega_1-\omega_2|$ and is zero if $\omega_1$ and $\omega_2$ have the same sign. Furthermore, we can use $\sigma_y \sigma^T_{i} \sigma_y= - \sigma_{i}$ (for $i=x,y,z$) in the two cooperon spin traces and obtain
\begin{eqnarray}\label{cooperon_final}
\overline{\chi_{ij}\chi_{i'j'}}({\bf R}) && = \frac{(m\tau)^4}{2\pi^2} \int_0^\infty \omega  d\omega \nonumber \\
\times  \sum_{\mu, \nu, \mu', \nu'}&& [ D^{\mu\nu}_{\bf R}(\omega)D^{\mu'\nu'}_{-\bf R}(\omega) + C^{\mu\nu}_{\bf R}(\omega)C^{\mu'\nu'}_{-\bf R}  (\omega) ] \nonumber \\
&&\times {\rm Tr}(\sigma_i \sigma_\mu \sigma_{i'}\sigma_{\nu'}){\rm Tr}(\sigma_j \sigma_{\mu'}\sigma_{j'}\sigma_{\nu}). 
\end{eqnarray}
Finally, as a consequence of time-reversal symmetry, $\sigma_y G_\omega^T(-{\bf k})\sigma_y = G_\omega({\bf k})$ which implies $X_D=X_C$ and $D_{\bf R}(\omega)=C_{ \bf R}(\omega)$.\cite{Chalaev2009} Therefore, the cooperon and diffuson contributions are equal and Eq.~(\ref{correlations_formula}) is obtained.

\section{Equal Rashba and Dresselhaus couplings}\label{app_ReqD}

The result for this case was discussed in Sec.~\ref{sec_equalRD} in terms of a simple unitary transformation. We show here that Eq.~(\ref{correlations_formula}) is in agreement with this exact argument, even if $X_D({\bf q},\omega)$ is approximated as in Eq. (\ref{X_D_result}). By using $U |{\bf k} \rangle= \frac12 \sum_\pm (1\pm \sigma_y)|{\bf k} \mp  m\alpha_+{\bf e}_x \rangle $ we can write
\begin{equation}
G_{\omega}({\bf k})= \frac12 \sum_{\pm} (1 \pm \sigma_y) G^{(0)}_\omega({\bf k} \mp m\alpha_+ {\bf e}_x  ),
\end{equation}
with $G^{(0)}_\omega({\bf k})$ the Green's function in the absence of spin-orbit coupling. Inserting this formula in Eq.~(\ref{XD_def}) it is easily obtained that, at $\alpha_-=0$,
\begin{eqnarray}\label{XD_transformd_wU}
&& X^{\mu\nu}_D({\bf q},\omega)=\frac18 \sum_{\pm} \{ {\rm Tr}[\sigma_\mu(1 \pm \sigma_y)\sigma_\nu(1 \pm \sigma_y)] X_D^{(0)}({\bf q},\omega)  \nonumber \\
&&+{\rm Tr}[\sigma_\mu(1 \pm \sigma_y)\sigma_\nu(1 \mp \sigma_y)] X_D^{(0)}({\bf q} \mp 2 m\alpha_+ {\bf e}_x,\omega) \}.
\end{eqnarray}
The value of $\overline{\chi_{ij}\chi_{i'j'}}$  without spin-orbit coupling of Eq.~(\ref{corr_chi_noSO}) was obtained by using $X_D^{(0)}({\bf q},\omega)\simeq 1-(q\ell)^2/2 -|\omega|\tau$. Substituting this expression in (\ref{XD_transformd_wU}) above, we find the same result of Eq.~(\ref{X_D_result}) at $\alpha_-=0$. Therefore, Eq.~(\ref{X_D_result}) satisfies the unitary transformation argument and direct numerical evaluation of Eq.~(\ref{correlations_formula}) has to be in agreement with the discussion of Sec.~\ref{sec_equalRD}.

\section{Details on the evaluation of $\overline{\chi_{ij}\chi_{i'j'}}$ with only Rashba coupling}\label{app_delta13}

To evaluate $\overline{\chi_{ij}\chi_{i'j'}}$ it is convenient to rescale in Eq.~(\ref{correlations_formula}) the frequency and the diffuson correlator as
\begin{equation}\label{w_def}
w= 2\omega \tau ( R/\ell)^2,  \qquad d_{\bf R}(w)= 2\pi m \tau \ell^2 D_{\bf R}(\omega).
\end{equation}
such that the tensor amplitude appearing in Eq.~(\ref{chichi_compact}) is given by
\begin{eqnarray}\label{eta_generaldef}
\eta_{ij,i'j'}=\frac{3}{32} \int_0^\infty w dw \sum_{\mu,\nu, \mu',\nu'} d^{\mu\nu}_{\bf R}(w)d^{\mu'\nu'}_{\bf R}(w) \nonumber \\
\times {\rm Tr}(\sigma_i \sigma_\mu\sigma_{i'}\sigma_{\mu'}){\rm Tr}(\sigma_j \sigma_{\nu'}\sigma_{j'}\sigma_{\nu}).
\end{eqnarray}
It is also convenient to use the dimensionless variable $\boldsymbol{\kappa} ={\bf q} R$ in the Fourier transform (\ref{diffuson_def})
\begin{eqnarray}\label{diffuson_def_rescaled}
d_{\bf R}(w)=  && \int \frac{\kappa d\kappa d\phi }{(2\pi)^2}  e^{-i \kappa \cos(\phi-\varphi)} \nonumber \\
&& \times \frac{\pi \ell^2}{R^2} \left[\mathds{1}-X_{D}\left(\boldsymbol{\kappa}/R, w(\ell/R)^2/2\tau \right)\right]^{-1}, \quad
\end{eqnarray} 
where it is found that the second line of Eq.~(\ref{diffuson_def_rescaled}) only depends on the two ratios $R/\lambda_\pm$. Therefore, also $\eta_{ij,i'j'}$ only depends on these two parameters, beside $\varphi$ (the azimuthal angle of $\bf R$). For example, in the case of Rashba spin-orbit coupling ($\beta=0$), we can assume $\varphi=0$ and perform the angular integration in terms of Bessel functions $J_i(\kappa)$ to obtain
\setlength\arraycolsep{2pt}
\begin{equation}\label{d_zeroes}
d_{\bf R}(w)= \int_0^\infty \kappa d \kappa  \left(
\begin{array}{cccc}
d^{00}_{\kappa}(w) & 0 & 0 & 0 \\
0 & d^{+}_{\kappa}(w) & 0 & d^{xz}_{\kappa}(w)  \\
0 & 0 & d^{-}_{\kappa} (w) & 0 \\
0 & -d^{xz}_{\kappa}(w)  & 0 & d^{zz}_{\kappa}(w)  \\
\end{array}
\right)~,
\end{equation}
with
\begin{eqnarray}
&& d^{00}_{\kappa}(w) =\frac{ J_0(\kappa)}{\kappa^2+w}~, \quad  d^{xz}_{\kappa}(w) =\frac{4 r \, \kappa J_{1}(\kappa)}{f(r,\kappa,w)}~,  \nonumber \\
&& d^{zz}_{\kappa}(w) =\frac{(\kappa^2+w+4r^2)J_{0}(\kappa)}{f(r,\kappa,w)}~, \nonumber\\
&& d^{\pm}_{\kappa}(w) =
\frac{[f(r,\kappa,w)+8 r^2\kappa^2]J_{0}(\kappa)\mp 8 r^2\kappa^2 J_{2}(\kappa)}
{f(r,\kappa,w)(\kappa^2+w+4r^2)}~, \nonumber
\end{eqnarray}
where $r=R/\lambda$ and
\begin{equation}
f(r,\kappa,w)=(\kappa^2+w)^2-4r^2(\kappa^2-3w)+32 r^4.
\end{equation}
To obtain $\eta_{ij,i'j'}$ from Eq.~(\ref{eta_generaldef}), the remaining integrations (in $d\kappa$ and $dw$) are evaluated numerically. We note that not all $\eta_{ij,i'j'}$ are independent. Some general symmetry transformations of $\eta_{ij,i'j'}$ are discussed in Appendix~\ref{eta_simmetries} but additional equalities exist in the case of Rashba spin-orbit coupling, due to the simplified structure of the diffuson matrix (\ref{d_zeroes}). For example, $\eta_{xx,xx}=\eta_{zz,zz}$ and $\eta_{xx,xz}=\eta_{xz,xx}$, as discussed in Sec.~\ref{sec_I12_both_perp}. Furthermore, several components are zero (in particular, $\eta_{iy,ij'}=0$ if $j'=x,z$). The independent components of $\eta_{ij,i'j'}$ are obtained
in terms of the integrals $\int_0^\infty  d_{\bf R}^{\mu\nu}d_{\bf R}^{\mu'\nu'} w dw$, for example $\eta_{yy,yy}=A_0$ is given by
\begin{equation}\label{nu_yyyy}
\eta_{yy,yy}=\frac{3}{8}\int_0^\infty \left\{ \sum_{i} [d_{\bf R}^{ii}(w)]^2+2[d_{\bf R}^{xz}(w)]^2\right\}wdw.
\end{equation}
where the sum runs over $i=0,x,y,z$. 

Due to their special relevance, we give below the expressions following from Eqs.~(\ref{Phi_def}), (\ref{Phi1_def}), and (\ref{Phi_pm_def}) for $\Phi$, $\Phi^\prime$ and $\Phi^\pm$:
\begin{eqnarray}
\label{tan_Phi_int}
& \tan 2\Phi & =\frac{2 \int_0^\infty 
d_{\bf R}^{xz} \left( d_{\bf R}^{zz} - d_{\bf R}^{xx} \right) w dw}{\int_0^\infty 
\left[(d_{\bf R}^{zz})^2 - (d_{\bf R}^{xx})^2 \right]  w dw }, \\
\label{tan_Phi1_int}
&\tan 4\Phi^\prime  & =\frac{4 \int_0^\infty  
d_{\bf R}^{xz} \left( d_{\bf R}^{xx} + d_{\bf R}^{zz} \right) w dw}{\int_0^\infty 
[(d_{\bf R}^{xx} + d_{\bf R}^{zz})^2 -4 (d_{\bf R}^{xz})^2] w dw }, \\\label{tan_Phip_int}
&\tan 2\Phi^+  & =\frac{ 2\int_0^\infty 
d_{\bf R}^{yy}  d_{\bf R}^{xz} w dw}{\int_0^\infty 
d_{\bf R}^{yy}(d_{\bf R}^{xx} + d_{\bf R}^{zz}) w dw }, \\\label{tan_Phim_int}
&\tan 2\Phi^-  & =\frac{ 2\int_0^\infty 
d_{\bf R}^{00}  d_{\bf R}^{xz} w dw}{\int_0^\infty 
d_{\bf R}^{00}(d_{\bf R}^{xx} + d_{\bf R}^{zz}) w dw }.
\end{eqnarray}
As seen from Eq.~(\ref{d_zeroes}), the diffuson matrix can be diagonalized at each value of $w$ with a rotation around ${\bf e}_y$ but the angle is determined by the ratio $d_{\bf R}^{xz} /(d_{\bf R}^{xx} + d_{\bf R}^{zz} )$ and in general depends on $w$. However, the results of Sec.~\ref{sec_rashba_only} imply that such ratio is approximately independent of $w$. By writing
\begin{equation}\label{lambda_def}
\frac{2 d_{\bf R}^{xz}}{d_{\bf R}^{xx} + d_{\bf R}^{zz} }  \simeq \tan \Phi ,
\end{equation}
where $\Phi$ is a function of $R/\lambda$ (but not of $w$), Eqs.~(\ref{tan_Phi_int})-(\ref{tan_Phim_int}) can be simplified to give $\Phi \simeq \Phi^\prime \simeq \Phi^\pm$.

As a final remark, we note that Eq.~(\ref{lambda_def}) also implies $A_0 \simeq A_0^\prime$ and therefore Eq.~(\ref{general_final}) for the variance. This is because the matrix $ d_{\bf R}(w)$ becomes approximately diagonal with new rotated frames for ${\bf I}_{1,2}$. By denoting as $\tilde{d}_{\bf R}(w)$ the transformed (diagonal) diffuson, we have
\begin{eqnarray}
& A_0 &=\frac38 \int_0^\infty  [(\tilde{d}^{00}_{\bf R})^2+(\tilde{d}^{xx}_{\bf R})^2+(\tilde{d}^{yy}_{\bf R})^2+(\tilde{d}^{zz}_{\bf R})^2]wdw,  \nonumber \\
& A_x &=\frac38 \int_0^\infty  [(\tilde{d}^{00}_{\bf R})^2+(\tilde{d}^{xx}_{\bf R})^2-(\tilde{d}^{yy}_{\bf R})^2-(\tilde{d}^{zz}_{\bf R})^2]wdw,  \nonumber \\
& B_x^- &=\frac34 \int_0^\infty \tilde{d}^{yy}_{\bf R} \tilde{d}^{zz}_{\bf R} wdw, \nonumber \\
& B_x^+ & =\frac34 \int_0^\infty \tilde{d}^{00}_{\bf R} \tilde{d}^{xx}_{\bf R} wdw, \nonumber
\end{eqnarray}
while the other coefficients of Eq.~(\ref{general_final}) are obtained from cyclic permutations of $xyz$. In particular, $A_0 = A_0^\prime$ within this approximation.

\section{Symmetries of $\eta_{ij,i'j'}$}\label{eta_simmetries}

One relation that the $\eta_{ij,i'j'}$ have to satisfy is simply due to $\overline{\chi_{ij}\chi_{i'j'}}=\overline{\chi_{i'j'}\chi_{ij}}$, which gives
\begin{equation}\label{symm_eta_1}
\eta_{ij,i'j'}({\bf R})=\eta_{i'j',ij}({\bf R}).
\end{equation}
A second relation can be obtained from relabeling of the magnetic moments 1 and 2
\begin{equation}\label{eta_transf_2}
\eta_{ij,i'j'}({\bf R})=\eta_{ji,j'i'}(-{\bf R}).
\end{equation}
Furthermore, the electron Hamiltonian is invariant upon a $\pi$ rotation around $z$ and we can transform the tensor $\eta_{ji,j'i'}(-{\bf R})$ in Eq.~(\ref{eta_transf_2}) to obtain
\begin{equation}\label{symm_eta_2}
\eta_{ij,i'j'}({\bf R})= \eta_{ji,j'i'}({\bf R}) \mathcal{R}_{ii}\mathcal{R}_{jj}\mathcal{R}_{i'i'}\mathcal{R}_{j'j'},
\end{equation}
where $\mathcal R_{ij}=\delta_{iz}\delta_{jz}-\delta_{ix}\delta_{jx}-\delta_{iy}\delta_{jy}$.


\end{document}